\documentclass[prd,amsmath,amsfonts,preprint,12pt,superscriptaddress]{revtex4}

\usepackage{amsmath}

\def\ba{\begin{eqnarray}}
\def\ea{\end{eqnarray}}
\def\be{\begin{equation}}
\def\ee{\end{equation}}

\newcommand{\qb}[1]{\Big\langle #1 \Big\rangle}
\newcommand{\er}[1]{(\ref{#1})}
\newcommand{\mc}[1]{\mathcal{ #1 }}
\newcommand{\vtxa}[1]{\mathcal{V}_{#1}}
\newcommand{\vtxb}[2]{\mathcal{V}_{#1}(#2)}
\newcommand{\vtxc}[3]{\mathcal{V}_{#1}^{#3}(#2)}
\newcommand{\saction}{\mathcal{S}_\phi}
\newcommand{\sactionp}[1]{\mathcal{S}^{(#1)}_\phi}
\newcommand{\h}[1]{h^{(#1)}_{\mu\nu}}

\begin{document}

\title{String Pair Production in a Time--Dependent Gravitational Field}
\author{Andrew J. Tolley}
\email{atolley@princeton.edu}
\affiliation{Joseph Henry Laboratories, Princeton University, Princeton NJ, 
08544}
\author{Daniel H. Wesley}
\email{dhwesley@princeton.edu}
\affiliation{Joseph Henry Laboratories, Princeton University, Princeton NJ, 
08544}

\date{September 29, 2005}

\begin{abstract}
\noindent We study the pair creation of point--particles and strings in a 
time--dependent,
weak gravitational field.  We find that, for massive string states, 
 there are surprising and
significant differences between the string and point--particle results.
Central to our approach is the fact that a weakly 
curved spacetime 
can be represented by
a coherent state of gravitons, and
therefore we employ 
standard techniques in string perturbation theory.
String and point--particle pairs are created through tree--level interactions
between the background gravitons.  In particular, we focus on the production of
excited string states and perform explicit calculations of the production
of a set of string states of arbitrary excitation level.  The differences
between the string and point--particle results may contain important 
lessons for the pair production of strings in the strong gravitational fields 
of interest in cosmology and black hole physics.

\end{abstract}

\maketitle
\newpage

\section{Introduction}\label{s:s1}

The aim of this paper is a calculation the pair production of strings in a
time--dependent background.  This phenomenon is of paramount importance
in cosmology, particularly during (p)reheating after inflation, and near
cosmological singularities.  Usually one studies pair production using
an effective field theory approach.  Here we study a spacetime with weak
gravitational fields and compute the pair production of excited string
states using worldsheet methods.  This ``stringy" calculation gives
significantly different results than those found in the effective field
theory approximation.  Our results suggest specific signatures of the
string pair creation phenomenon that may survive into the regime of
strong gravitational fields.
 
Recently 
\cite{Gubser:2003vk,Gubser:2003hm,Friess:2004zk} the pair creation of 
strings in a cosmological setting was 
estimated using effective field theory methods. 
In this work the excited states of
the string are treated as massive particles for which the usual field
theory calculation applies.  While the production rate of a single
massive state is small, the exponential Hagedorn growth of the number of
string states can give rise to a significant cumulative effect.
Here we test this effective field theory approximation in a weak--field limit,
where string pair creation may be computed exactly. 
Although our real interest is 
strong gravitational fields, such as those that arise in cosmology, 
the formalism does not yet exist to calculate pair production of 
strings here (see \cite{Lawrence:1995ct} for an interesting approach to 
this and \cite{Turok:2004gb} for a calculation in a more specific background).  Nevertheless interesting surprises can be seen already for weak 
 fields. Among them, we find that the enhancement of the production of excited string 
states due to the growing density of states is greatly suppressed due 
to the very mild hard scattering behavior of the string, which in turn arises from the fact that the string is an extended object.  

To carry out our calculation of string pair production, we use
the familiar idea that 
a curved geometry can be represented as a coherent state of 
gravitons \cite{Pol98,GSW87}. For weak gravitational 
fields this gives a prescription relating $S$-matrix amplitudes 
on a weakly curved spacetime to a sum of those on Minkowski spacetime with multiple 
graviton vertex operators inserted.  
To lowest order in the closed string coupling, the pair production of
strings is thus the four--point process $gg \to AB$, with two gravitons
$g$ interacting to produce two strings in states $A$ and $B$.
This approach has the advantage of leading to a controlled calculation
using standard string perturbation theory, but there are many other conceptual
and technical 
obstacles to an understanding of strings in general time--dependent backgrounds.
There are subtleties involved in interpreting the $S$--matrix for strings
on curved backgrounds (see, for example,
\cite{Hughes:1988bw,Campbell:1990dz}).
More generally, in any quantum theory including gravity,
the inevitable formation of singularities as predicted by the
singularity theorems \cite{Hawking73,Hawking:1976ra} implies the
absence of asymptotically 
trivial ``in" and ``out" regions required to define an $S$--matrix.
We construct a specific geometry describing the collision of two small amplitude plane waves which we embed in an exact nonlinear solution of Einstein's equations. An analysis of the full solution shows the formation of a spacelike singularity in the future of the collision of the plane waves, and also how by making the amplitude of both waves sufficiently small we may push the singularity arbitrarily far into the future of the initial collision region. 
Consequently perturbation theory is valid for an arbitrarily long period, and we believe the 
$S$--matrix formalism can be used to understand the physics in this region. This 
is as it should be, for the possible formation of a big crunch in our future should 
not prevent us from understanding physics today! 

In this work we primarily focus on the pair production of long--wavelength string
states.  Since the theory of pair creation in field theory is well developed
\cite{DeWitt:1975ys,Bir84}, and since string theory should have an effective
field theory description valid for $\lambda \gg \sqrt{\alpha'}$, one might ask
why we expect there to be differences between string and point particle pair creation in this regime.  When viewed as the four--point process $gg\to AB$, then
certainly for massless states $A$ and $B$,
with wavelengths $\lambda\gg\sqrt{\alpha'}$, all momentum scales in the scattering
process are well below the string scale.
On the other hand, to produce excited
string states, with masses $m \sim 1/\sqrt{\alpha'}$, then the typical
graviton momenta and momentum exchange will in fact be greater than 
$1/\sqrt{\alpha'}$.  This is beyond the validity of effective field theory,
and indeed previous investigations have revealed 
 uniquely ``stringy" behavior in this 
regime \cite{Gross:1987ar,Gross:1987kz,Gross:1988ue}.
Thus, it is reasonable to expect the pair
production of massive \emph{strings} might be fundamentally different from the pair
production of massive \emph{particles}.

Our calculation reveals two specific differences between the string and 
point--particle results.  As we have mentioned above, and will discuss in more 
detail in Section \ref{s:s5},  the pair production of
strings is suppressed due to the 
mild hard scattering behavior of string
amplitudes.  Another intriguing difference relates to the production of identical
string states.  At tree level in field theory only identical particle pairs 
are produced.
A surprising feature of the string case is that production of pairs of strings
at the same excitation level are suppressed, going to zero as
$\lambda \to \infty$.  Instead, most created pairs are of different mass levels.
Both of these features can be explained in the context of certain simple features
of the string spectrum and scattering amplitudes, as we discuss in more detail
below.

This paper is organised as follows: In Section \ref{s:s2} we discuss the 
perturbative approach to calculating pair production, using field theory as a guide 
to the string calculation. In Section \ref{s:s3} we discuss a specific example of a 
background, describing the collision of two plane waves, for which the 
perturbative calculation may be applied. 

In Section \ref{s:s5} we perform the explicit string 
calculation and discuss its properties.  Finally, we present our conclusions in 
Section \ref{s:s6}.  We also include two appendices with additional
details on some elements of our work. The four--point amplitude
describing pair production is discussed in Appendix \ref{appendix}, and 
a more detailed account of perturbatively computing Bogolubov coefficients
is given in Appendix \ref{appendix2}.

\section{Pair Production in Weak Gravitational Fields}\label{s:s2}

In this section we discuss how pair production may be studied
perturbatively.  Certainly there are examples where the production rate is
nonperturbative; these include Schwinger's classic calculation of 
$e^+ e^-$ production in an electric field \cite{Schwinger:1951nm}
and more recent calculations with open strings
\cite{Bachas:1992bh,Burgess:1986dw,Cho:2005aj}.  An example with a different
behavior is provided by 
the pair production of scalars of mass $m$ 
in a flat FRW spacetime that undergoes a brief 
period of expansion \cite{Bernard:1977pq}.  
In this example we have the metric
\begin{equation}
ds^2 = a(\eta)^2 \left( -d\eta^2 + dx_3^2 \right), \qquad
a^2(\eta) = 1 + b \tanh (\rho \eta ),
\end{equation}
and the number of particles produced in each mode
$k = (\omega, \vec k)$ is found to be
\begin{equation}\label{eq:FRW_step}
N_{\vec k} = \frac{\sinh^2 \left[ \pi (\omega_{+\infty} - \omega_{-\infty})
/ 2\rho \right]}{\sinh (\pi \omega_{+\infty}/\rho) \sinh 
(\pi \omega_{-\infty}/\rho)}
\end{equation}
where
\begin{equation}
\omega(\eta) = \left[ |\vec k|^2 + a(\eta)^2 m^2 \right]^{1/2}, \qquad
\omega_{\pm \infty} = \omega (\pm \infty).
\end{equation}
For $b  \ll 1$ the
number of pairs produced goes like $\sim b^2m^4$ for small $b$.  
Thus there are backgrounds in which is is plausable that pair creation could
be studied within a perturbative framework.

\subsection{Field Theory}

In the case of field theory, studying pair production can be 
realized quite straightforwardly through the addition of new vertices
in the Feynman diagrams of the theory.  This point of view is developed in
more detail in Appendix \ref{appendix2}, where it is shown that this
technique yields precisely the same results as the more standard Bogolubov
coefficient calculation.  The field theory case is useful for our present
purposes as it provides a  prototype for the string calculation.
One expands the metric about Minkowski space
\begin{equation}\label{eq:qft_metric}
g_{\mu\nu} = \eta_{\mu\nu} + \h{1} + \h{2} + \cdots,
\end{equation}
where $\h 1$ is the first order metric perturbation, and
$\h 2+ \dots$, the corrections required at successively higher
order so that the full metric $g_{\mu\nu}$ satisfies the Einstein equations.
If we consider a massive minimally coupled scalar field with
action
\begin{equation}\label{eq:qft_action}
\saction = -\frac{1}{2} \int \left(
g^{\mu \nu} \partial_\mu \phi \partial_\nu \phi
+ m^2 \phi^2  \right) \sqrt{-g}  \; d^D x ,
\end{equation}
Then the expansion \er{eq:qft_metric} of the metric 
gives a corresponding perturbative expansion of the action,
\begin{equation}\label{eq:qft_action_ex}
\saction = \sactionp{0} + \sactionp{1} + \sactionp{2} + \cdots
\end{equation}
as well as a  series of terms $\sactionp n$ at $n^{th}$ order in the
perturbation.

This allows a perturbative evaluation of the $n$--point functions 
in the background, denoted by $\langle \cdots \rangle_{b}$, in terms
of the same quantities $\langle \cdots \rangle_{0}$ in Minkowski space with
additional insertions.  For any combination of fields $\mathcal{O}$,
the expansion  of the action \er{eq:qft_action_ex} inserted into the
path integral yields,  
\begin{equation}\label{eq:qft_2pt_ex}
\langle \mathcal{O} \rangle_b =
\langle \mathcal{O} \rangle_0 + 
\int
\langle T\mathcal{O}_k \Sigma^{(1)}_{-k} \rangle_0 
\frac{d^D k}{(2\pi)^D} + 
\int 
\langle T\mathcal{O}_{k'} \Sigma^{(2)}_{-k'} \rangle_0 
\frac{d^D k'}{(2\pi)^D}+ \cdots
\end{equation}
where $\Sigma^{(1)}$ and $\Sigma^{(2)}$ are 
the new vertices that appear to first and second order
in perturbations.  
These are determined by the expansion \er{eq:qft_action_ex} of the action
\begin{subequations}
\begin{align}
 \Sigma^{(1)} & = i \sactionp 1, \\
 \Sigma^{(2)} & = i \sactionp 2 - \frac{1}{2} \left[ \sactionp 1 \right]^2.
\end{align}
\end{subequations}
The statements that we have made for the $n$--point functions have analogues for
the $S$--matrix.  However, defining
an $S$--matrix involves a choice of asymptotic ``in" and ``out" states. 
Therefore this prescription really only makes sense when the spacetime in question
is asymptotically Minkowski.  In other cases it may be necessary to modify the
definition of asymptotic states.  

At this point, it is possible to see how even a weak gravitational
background can lead to pair production.  
Pair production is the process with no ``in" particles and two ``out" particles.
While forbidden by momentum conservation in Minkowski space,
in the perturbed spacetime each term in the $S$--matrix carries three
momenta; the two momenta of the ``out" particles and 
the momentum carried by the new vertices $\Sigma^{(n)}$.
Thus two positive energy particles can appear in the ``out" state if 
one of the terms  $\Sigma^{(n)}_{k}$ supplies the ``missing" momentum.  
So, the amplitude to pair produce two particles with momenta
$k_1, k_2$ will be nonzero provided that we have 
$\Sigma^{(n)}_{-k_1 - k_2} \ne 0$. More physically the missing momentum is 
coming from the background gravitons.  

\subsection{String Theory}

Following the same procedure as the field theory calculation, one may
insert the perturbative expansion of the metric 
$g_{\mu\nu} = \eta_{\mu\nu} + \h 1 + \h 2 + \cdots$
into the conformal gauge string worldsheet action
\begin{equation}\label{eq:ws_action}
S = - \frac{1}{2\pi\alpha'} \int g_{\mu\nu} (X)
\partial X^\mu {\bar \partial} X^\nu
\; d^2 z .
\end{equation}
Now, the successive metric perturbations $\h n$ are chosen to satisfy the
string beta function equations.

As is well known \cite{GSW87,Pol98} 
the expansion of the metric leads to the insertion of
graviton vertex operators in worldsheet correlation functions.
For example, if we wish to study the $S$--matrix element involving two
string states $A$ and $B$ then we must compute the worldsheet correlation
function
\begin{equation}\label{eq:str_1st_order}
\qb{ \vtxb{A}{k_A} \vtxb{B}{k_B} }_b = 
\qb{ \vtxb{A}{k_A} \vtxb{B}{k_B} }_0 +
\frac{1}{4\pi g_c}\int h^{(1)}_{\mu\nu}(k')
\qb{ \vtxb{A}{k_A} \vtxb{B}{k_B} \vtxc{g}{k ' }{\mu\nu}} _0 
\; \frac{d^D k'}{(2\pi)^D}  \\ + \cdots
\end{equation}
where $\vtxb{A,B}{k_{A,B}}$ are the vertex operators for the created
strings in states $A$ and $B$, $\vtxc{g}{k'}{\mu\nu}$ is a graviton
vertex operator integrated over the worldsheet representing the metric perturbation, and
we have absorbed a factor of $g_c$ in each of the vertex operators. The 
metric perturbation $\h 1$ appears as 
the polarization tensor of the background graviton.  

The first--order contribution to pair creation \er{eq:str_1st_order}
vanishes when $A$ and $B$ are massive, due to momentum conservation
when the graviton is on--shell (as it must be for a consistent string
amplitude).  Nevertheless, \er{eq:str_1st_order} is related to the
phenomenon of ``particle transmutation," by which a string can change its
mass and spin as it passes through a gravitational background
\cite{Horowitz:1990sr, deVega:1990cu, DeVega:1992wf, Tolley:2005us}.  We will not 
investigate this phenomenon in the current work, but
for weakly curved spacetimes it could in principle be
studied within the $S$--matrix framework we apply herein.

Ideally we would be able to continue this expansion to higher order in
the metric perturbation, by analogy  
to the field theory case.  However, at second order issues arise in the
string theory that make the situation somewhat more complex.
For one, consistency of the of the string amplitude (\ref{eq:str_1st_order})
requires that each of
the vertex operators appearing in be a conformal tensor of weight 
$(h, \bar h) = (1,1)$, or in other words satisfy the physical state
conditions. As is well known, this is
equivalent to imposing the linearized 
Einstein equations on $h^{(1)}_{\mu\nu}$. 
At second order in the expansion of worldsheet amplitudes, a graviton vertex 
operator with polarization
$\h 2$  appears.  By construction, the second--order perturbation
$\h 2$ satisfies the second--order RG equations.
In general, this means that $\h 2$ with \emph{not} in fact satisfy the physical
state conditions.
Thus, the
string amplitude with
the corresponding vertex operator will not be consistent.  

Another issue that arises is the presence of a nonzero ``tadpole"
for some fields.  Taking our prescription literally, to compute the
$S$--matrix elements on our background to second order involves inserting
two graviton vertex operators into the relevant correlation functions.
For massive fields $A$, the tadpole is zero at first order due to 
momentum conservation.
At second order in the metric perturbation, the tadpole
for some state $A$ in the background is therefore given by
\be
\qb{ \vtxb{A}{k_A} }_b = 
\qb{ \vtxb{A}{k_A} \vtxb{g}{k_1} \vtxb{g}{k_2} } + \cdots
\ee
with $\vtxb{g}{k} = \frac{1}{4\pi g_c} \h 1 \vtxc{g}{k}{\mu\nu}$.  This expression
does not vanish for the examples we consider.
A nonvanishing tadpole is an indication that the
classical equations of motion are not exactly obeyed by the background 
solution.  This is to be expected, as here 
we will use a background solution that solves the leading--order
equations of motion for massless string excitations, but we have
ignored the massive excitations and higher corrections.  
The presence of the tadpole is telling us 
that the equations of motion for these massive fields forbid
setting them to zero throughout spacetime.
Furthermore, in a curved background the vertex operators 
and states should change, which we have not taken into account.
We expect that the failure of the tadpole to vanish is due to some
combination of all of these effects.

Given these new complexities, it
appears that problems are beginning to develop with
the technique of inserting new vertex operators at second order.  
Nevertheless, by considering each of these new problems, we will
argue in Sections \ref{coherent} that useful results can be obtained from the
second order expansion.  Therefore, for the purpose of studying pair production
we will use the prescription
\be \label{eq:str_prescription}
\langle AB | S | vac \rangle = 
\frac{1}{(4\pi g_c)^2} \int h^{(1)}_{\mu\nu}(k') h^{(1)}_{\sigma\tau}(k'')
\qb{ \vtxb{A}{k_A} \vtxb{B}{k_B} \vtxc{g}{k ' }{\mu\nu}
\vtxc{g}{k '' }{\sigma\tau}
}_0  \frac{d^Dk'}{(2\pi)^D}\frac{d^Dk''}{(2\pi)^D}
\ee
where $\vtxb{A,B}{k_{A,B}}$ are the vertex operators for the pair created
string states $A$ and $B$.  In defining this S-matrix element it is understood as usual that the positions of three of the vertex operators are fixed and the remaining one is integrated over the worldsheet. This prescription does not involve a term at
first order in the metric perturbation, since the pair production of massive
states is forbidden to this order by momentum conservation.  Furthermore
corrections to the vertex operators and background solution, which should
be included in a more complete analysis, only affect this 
$S$--matrix element at higher orders in $\h 1$.  

\subsection{Coherent states}
\label{coherent}

The existence of non--zero ``tadpoles," or 3--point functions
$\langle \vtxa g \vtxa g \vtxa N \rangle $
where $\vtxa N$ denote the vertex operators for level $N$ string states, 
implies that 
turning on a nontrivial gravitational background will, at second order, 
produce a nontrivial background of excited states. 
The question we must address is whether we can still trust the pair production
results that we will extract at this order.  By studying an analogous situation
in field theory, we will argue that the corrections to the pair production
amplitudes appear at higher order, and thus our second order results may be trusted.

The production of excited string states can be modeled in field theory as
 the pair production of a 
massive scalar field $\chi$ via interactions with a massless scalar field 
$H$, representing the graviton. This can be realized with the action
\ba
S=\int d^d x \left( -\frac{1}{2} (\partial H)^2 -\frac{1}{2} (\partial \chi)^2 -\frac{1}{2}  m^2 \chi^2  +\lambda H^2 \chi -\frac{1}{2}\mu H^2\chi^2\right) .
\ea
The presence of the $\lambda$ coupling 
implies that we cannot turn on a background field for $H$ without 
also exciting a background value for $\chi$, as a consequence of the equations
 of  motion
\ba
-\Box H=2\lambda H \chi +\dots\\
-\Box \chi+m^2 \chi=\lambda H^2+\dots. .
\ea
The $\mu$ coupling gives rise to pair creation of $\chi$ quanta in the presence of a nontrivial `graviton' background $H$. Before calculating this effect we must first correctly deal with the background. Assuming $\chi=0+O(H^2)$ these equations may be solved perturbatively to second order in $H$ as (here $\int d\tilde{k}=\int d^{d-1}k/(2\omega)$, $\omega=|k^0|$ which is valid for massless and massive particles)
\ba
H(x)&=&H_0(x)=\int d\tilde{k} \left( \alpha^*(k) e^{ik.x}+\alpha(k)e^{-ik.x} \right)+O(H_0^3), \\
\chi(x)&=&\lambda \int d^dy G_{ret}(x,y) H^2_0(y)+O(H_0^3),
\ea
where $G_{ret}(x,y)$ is the retarded propagator for a scalar of mass $m$ and we have chosen the following boundary conditions
\be
\lim_{t \rightarrow -\infty} \chi(x) =0.
\ee
It is well known that in field theory a nontrivial background may be described by a coherent state. The initial state contains only gravitons and may be described by 
\be
|vac,t \rightarrow -\infty \rangle = e^{-\frac{1}{2}\int d\tilde{k} |\alpha|^2} e^{\int d\tilde{k}\alpha a^{\dagger}}|0 \rangle.
\ee
Whereas the final state is a coherent state of both `gravitons' and $\chi$ quanta
\be
|vac,t \rightarrow +\infty \rangle=e^{-\frac{1}{2}\int d\tilde{k}  |\beta|^2}e^{-\int d\tilde{k} \frac{1}{2}|\alpha|^2} e^{\int d\tilde{k} \alpha a^{\dagger}}e^{\int d\tilde{k}\beta d^{\dagger}}|0 \rangle.
\ee
Here $a^{\dagger}/a$, $d^{\dagger}/d$ are the conventionally defined creation/annihilation operators for $H$ and $\chi$ respectively, and $\beta(k)$ is the generated background for $\chi$
\be
\lim_{t \rightarrow +\infty} \chi(x) =\int d\tilde{k} \left( \beta^*(k) e^{ik.x}+\beta(k)e^{-ik.x} \right).
\ee
We now compute the amplitude to create a pair of $\chi$ quanta. Because of the nontrivial background for $\chi$ as $t \rightarrow +\infty$ we must redefine the creation/annihilation operators as $\tilde{d}=d-\beta$. 
This is the analogue of cancelling the tadpole in the string theory.
The $\beta$ are also determined by the condition that the amplitude to create
a single particle must vanish (tadpole cancellation)
\be
\langle vac,t \rightarrow +\infty|\sqrt{2\omega}\tilde{d}S|vac, t\rightarrow -\infty \rangle=0.
\ee
Now the amplitude to create a single pair of $\chi$ quanta with momenta $(k_3,k_4)$ is
\be
A[vac \rightarrow 2]=\langle vac,t \rightarrow +\infty|\sqrt{2\omega_3}\tilde{d}(k_3)\sqrt{2\omega_4}\tilde{d}(k_4) S|vac, t\rightarrow -\infty \rangle,
\ee
where $S$ is the $S$-matrix. 
Utilizing the fact that the amplitude to create a single particle must vanish
and expanding $A[vac \rightarrow 2]$ to second order we find after making uses of conservation of energy and momentum
\be
A[vac \rightarrow 2]=\int d\tilde{k_1} \int d\tilde{k_2} \, \, \alpha(k_1) \alpha(k_2) \langle 0|\sqrt{2\omega_3}d(k_3) \sqrt{2\omega_4} d(k_4)Sa^{\dagger}(k_1) a^{\dagger}(k_2) |0 \rangle +O(H_0^3).
\ee
This is a very simple result: to second order in perturbations, the amplitude to create a pair of particles is determined by the amplitude for two gravitons to convert to two $\chi$ quanta. 
Furthermore, at second order the result in independent of the modifications
to the creation operators required to cancel the tadpole.  This suggests that
our string results will be reliable, even though we will not carry out the
modifications to the vertex operators required to cancel the tadpole in the
string case.

\section{Background}\label{s:s3}

In this section we shall construct a specific example of a perturbative 
geometry for which our approach may be applied. In doing so we shall also 
uncover some of the inevitable limitations of the perturbative approach. The  
issue we must address is whether it makes sense to define a perturbative 
$S$-matrix on a time-dependent spacetime, due to the inevitable formation 
of spacelike singularities or black holes, as
predicted by the singularity theorems \cite{Hawking73,Hawking:1976ra}.

\subsection{Colliding plane waves}

One of the simplest and best studied time--dependent geometries is the pp-wave. 
This is a solution of Einstein's vacuum equations with metric given by
\begin{equation}
ds^2=2dudv+H(u,X) du^2+dX^idX_i,
\end{equation}
where $H(u,X)$ is a harmonic function $\nabla^2 H(u,X)=0$. These solutions may 
easily be extended to include nontrivial dilaton and NS 2-form sources. However,
 for simplicitly of presentation we shall concentrate on pure vacuum solutions.
We will focus on the class of ``exact" plane waves
for which $H(u,X)=A_{ij}(u)X^iX^j$ with $\sum_i A_{ii}(u)=0$. All pp-wave 
spacetimes admit a null Killing vector $\frac{\partial}{\partial{v}}$ and this 
allows a global definition of null time. As a consequence, the natural ``in" 
and ``out" field theory vacua are chosen by decomposing modes into positive 
and negative frequency with respect to this time, and since this is a global 
definition we find that for a free field $ |0\rangle_{in}=|0 \rangle_{out}$ 
implying the absence of particle creation \cite{Gibbons:1975jb}. (This is true provided we make the conventional choice of vacua. If 
the plane wave geometry is not asymptotically Minkowski at past and future 
infinity then the reasons for this choice are less clear). 

In order to obtain a background with pair creation we will consider a spacetime describing the collision of two plane waves.
Treating the 
amplitude of each wave $A^{(\alpha)}_{ij}$ as small, then at the linearized 
level this will be described by the metric,
\be
\label{metric1}
ds^2=2dudv+A^{+}_{ij}(u)X^iX^j du^2+A^{-}_{ij}(v)X^iX^j dv^2+dX^idX_i +O(A^2).
\ee
At the linearised level, the two waves pass through each other undisturbed. 
The $O(A^2)$ terms will include interactions between the waves. If the 
amplitude of each wave is localized around a given null time, for example if we
take
\be
A_{ij}^{+}(u) \approx A\exp{(-u^2/L^2)}
\ee 
then at leading order the interaction region will be localized near $(|u|<L, |v|<L)$. However, at higher orders we will see that there is a long tail in the future lightcone of the interaction region in which perturbations grow, eventually leading to a singularity. 

Fortunately all this can be seen explicitly in a concrete example where 
$A^{+}_{ij}$ and $A^{-}_{ij}$ are both diagonal. In this case it is 
straightforward to lift the linearised metric (\ref{metric1}) to an exact 
nonlinear solution of Einstein's equations. To see this, first perform a 
linearized gauge transformation to take the metric to the form
\be
\label{metric2}
ds^2=2dudv+(\delta_{ij}+h_{ij}(u,v))dX^idX^j +O(A^2).
\ee
where 
\be
h_{ij}(u,v)=2 \int \int A^{+}_{ij}(u) dudu+2 \int \int A^{-}_{ij}(v) dvdv. 
\ee
This metric is a special case of a more general class of solutions given by
\be 
ds^2=2e^{\Gamma(u,v)} dudv+\alpha(u,v) \sum_{i} e^{2\beta_i(u,v)}dx_i^2,
\ee
where $\sum_i \beta_i =0$. This is the form of an exact set of solutions to Einstein's equations known as Einstein-Rosen waves. They satisfy $\partial_u \partial_v \alpha =0$ and $\partial_u (\alpha \partial_v \beta_i) +\partial_u(\alpha \partial_v \beta_i)=0$. This form is invariant under transformations $u \rightarrow f(\bar{u})$, $v \rightarrow g(\bar{v})$ and $\exp(\Gamma) \rightarrow \exp(\bar{\Gamma})/(f_{,\bar{u}}g_{,\bar{v}})$. We may use this freedom to set $\alpha=(\bar{u}+\bar{v})\sqrt{2}=t$ and $z=(\bar{u}-\bar{v})\sqrt{2}$ then the $\beta_i$ satisfy
\be
\frac{1}{t}\partial_t(t\partial_t \beta_i)-\partial_z^2 \beta_i=0.
\ee
It is straightforward to see that as $t \rightarrow 0$ the solutions to this equation behave as $\beta_i \rightarrow c_i(z)+d_i(z) \ln t$. The resulting solution describes a Kasner metric where the Kasner exponents are functions of $z$, consequently the generic collision of two plane waves will result in an anisotropic singularity.

The existence of a singularity should cause us no surprise, it arises as a 
simple consequence of the singularity theorems. Although the singularity 
theorems do not strictly apply here as we have no matter, compactifying the 
geometry on one direction gives us a theory with gravity and a scalar field 
which does satisfy the singularity criteria. In the present context it implies 
that perturbation theory will inevitably breakdown in a finite time period. In 
the traditional definition of an $S$-matrix theory, the ``out" state is defined 
at future infinity, thus it would seem at first sight impossible to define an
 $S$-matrix on a spacetime with a singularity at future infinity. 

\subsection{Perturbative solution}

Our approach to dealing with this fundamental issue is a pragmatic one, for
if there 
exists a long period of time for which perturbation theory is valid, then the 
$S$-matrix should describe physics in that period. 
Let us now construct the perturbative solution to Einstein's equations. 
This will enable us to check that
nonlinear corrections remain small at low orders in the perturbation 
expansion, and to explore
the breakdown of perturbation theory.
To be consistent with the metric (\ref{metric2}), we take 
$\beta^i \sim O(A)$, $\alpha \sim 1+O(A^2)$ and $\Gamma \sim O(A^2)$. The 
perturbed metric up to and including 3rd order in perturbations is,
\ba
\alpha&=&1+A^2 (\alpha_+(u)+\alpha_-(v))+O(A^4) ,\\
\beta^i&=&A(\beta^i_+(u)+\beta^i_-(v))-\frac{1}{2} A^3 (\alpha_+(u) \beta^i_-(v)+\alpha_-(v) \beta^i_+(u))+O(A^4), \\
\Gamma&=&-\frac{1}{4}A^2 \sum_i \beta^i_{+}(u)\beta^i_{-}(v)+O(A^4),
\ea
where,
\ba
\label{cond1}
\nonumber
\alpha_+''(u)&=&-\frac{1}{4} \sum_i \beta^{i \, 2}_{+,u}, \\
\alpha_-''(v)&=&-\frac{1}{4} \sum_i \beta^{i \, 2}_{-,v} .
\ea
We are assuming that $\beta_+(u)$ and $\beta_-(v)$ are localized functions, 
for instance with a Gaussian profile $\beta_{+}=\exp(-u^2/L^2)$ and 
$\beta_{-}=\exp(-u^2/L^2)$. From this it is clear that up to 3rd order in $A$, 
$\beta^i$ and $\Gamma$ are both bounded functions localized near the interaction
region $u=v=0$. 
However, as a consequence of (\ref{cond1}), $\alpha$ is not bounded.  We choose
the solution
\be
\alpha_+(u)=-\frac{1}{4} \int_{-\infty}^u du_1 \int_{-\infty}^{u_1} \sum_i \beta^{i \, 2}_{+,u_2}(u_2),
\ee
and similarly for $\alpha_-(v)$.
From this it is clear that as $u \rightarrow +\infty$, then
$\alpha_+(u) \rightarrow C_++D_+ u,$
where,
\be
D_+=-\frac{1}{4} \int_{-\infty}^{\infty} du \sum_i \beta^{i \, 2}_{+,u} \neq 0.
\ee
Trying to remove this divergence by adding the homogeneous solution $-D_+ u$
causes $\alpha_-(u)$ to diverge as $u \rightarrow -\infty$. This is the first 
indication that perturbation theory breaks down at late times.  On dimensional 
grounds, $D_+ \sim O(1)/L$ and so this coordinate system breaks down at 
$u \approx L/A^2$, $v \approx L/A^2$. Evidently by making $A$ sufficiently 
small, we can push this region arbitrarily far into the causal future of the
 interaction region. 

Remarkably, to 3rd order in $A$, the growth of $\alpha$ is a coordinate 
artifact and does not reflect a genuine breakdown of perturbation theory. One 
may show this by removing the linear growth in $u$ by means of the following 
coordinate transformation,
\ba
x^i &\rightarrow& x^i (1+\frac{A^2}{2} (D_+ u+D_- v))+O(A^3), \\
u &\rightarrow& u+\frac{A^2}{4} D_+ \vec{x}^2+O(A^3), \\
v &\rightarrow& v+\frac{A^2}{4} D_- \vec{x}^2+O(A^3) .
\ea
Similarly all components of the Riemann tensor are finite and 
supported only near
$u=v=0$.
In fact we must go to 4th order in 
perturbations to see the first signal that perturbation theory is breaking down.
 An explicit calculation shows that certain components of the Riemann 
 tensor diverge linearly for large $u$ and $v$  in any orthonormal frame. 
 However, unsuprisingly this only occurs when 
 $u \approx L/A^2$, $v \approx L/A^2$. So again we may push the inevitable 
 breakdown of perturbation theory off to arbitrarily far in the future of the 
 region of interest. 

In what follows we shall only compute the string amplitudes to second order in 
the background perturbations. As we have seen, at second order the collision of 
two plane waves results in a localized interaction region and no pathologies 
occur. Thus there will be no problem interpreting the implications of our 
amplitudes. If we continue the amplitude calculations to higher order in 
perturbations, we may expect to see mildly pathological behaviour associated 
with the eventual breakdown in perturbation theory. However, it should always 
be possible to separate this from the information which describes the 
interaction region. These above observations show the fundamental limitations 
of the application of the $S$-matrix formalism to a time-dependent geometry, 
these issues would not arise in a more Schrodinger like prescription where we 
consider the state at a given time, rather than the transition amplitude to a 
state at future infinity. This suggests that string field theory or a similar 
formalism may be a more appropriate way to consider time-dependent spacetimes.

\section{Pair Production}\label{s:s5}

The first nonvanishing contribution to string pair production arises from
the four--point amplitude with two incoming gravitons. In this section we
discuss this amplitude for an explicit set of string states, 
and obtain results in
accord with previous investigations on the high--energy behavior of 
string scattering amplitudes \cite{Gross:1987ar,Gross:1987kz,Gross:1988ue}.
This suggests that the behavior we observe here may hold for more general
excited string states as well.  A detailed calculation of the
relevant amplitude may be found in  Appendix \ref{appendix}.

We calculate the pair creation of ``representative" string states
given in oscillator
notation by
\begin{equation}\label{eq:rep_state}
| \epsilon ; k \rangle = 
(N!)^{-1}
\epsilon_{\mu_1 {\bar \mu}_1 \dots \mu_N {\bar \mu}_N}
\left( \prod_{j=1}^N \alpha_{-1}^{\mu_j} {\tilde \alpha}_{-1}^{{\bar \mu}_j}
\right)
 |0 ; k\rangle ,
\end{equation}
where we assume that $N > 1$.
This set of states include a unique scalar state at each exciation level $N$, 
which provides the most direct comparison with field theory results.  

Clearly the polarization $\epsilon$
must be symmetric under the interchange of two holomorphic indices.
The physical state conditions are satisfied provided that
\begin{subequations}
\begin{align}
m^2 & = \frac{4}{\alpha'} (N-1). \\
k^{\mu_j} \epsilon_{\mu_1 {\bar \mu}_1 \dots \mu_j \dots \mu_N {\bar \mu}_N} 
&= 0, \qquad \mbox{for all } j. \\
\eta^{\mu_j \mu_k} \epsilon_{\dots \mu_j \dots \mu_k \dots} &= 0, 
\qquad \mbox{for all } j,k. 
\end{align}
\end{subequations}
along with similar conditions for the barred indices.
These states will have unit norm
provided that
\begin{equation}
\langle \epsilon ; k | \epsilon ; k \rangle = 1 \qquad \mbox{if} \qquad
\epsilon^{\mu_1 {\bar \mu}_1 \dots \mu_j \dots \mu_N {\bar \mu}_N} 
\epsilon_{\mu_1 {\bar \mu}_1 \dots \mu_j \dots \mu_N {\bar \mu}_N} = 1.
\end{equation}
The vertex operators corresponding to these states are given by
\begin{equation}\label{eq:rep_vtx}
\vtxb{N}{k} = g_c \epsilon_{\mu_1 {\bar \mu}_1 \dots \mu_N {\bar \mu}_N}
\left(
\frac{2}{\alpha'} \right)^N
: \left[
\prod_{j=1}^N \partial X^{\mu_j} {\bar \partial} X^{{\bar\mu}_j} \right]
e^{i k \cdot X}
(z, \bar z) : 
\end{equation}
where we have included a factor of the closed string coupling $g_c$ as is
conventional.

As we are interested in comparing string results to those in field theory,
we focus on the production of ``long wavelength" string states, or
equivalently those whose spatial momenta are small in comparison with
$1/\sqrt{\alpha'}$.  Since we are also considering the production of very 
massive states, the relevant process is therefore one in which
the pair of strings are created nearly on threshold.  In order to fix
notation, we take $k_{\pm}$ to be the momenta of the created strings, and
$-k_{L,R}$ the momenta of the incoming gravitons, so that the 
equation for conservation
of momentum is
\begin{equation}
k_+ + k_- + k_L + k_R = 0.
\end{equation}
We further take the created strings to be scalar representative states of
excitation levels $N_{\pm}$.  
For simplicity of exposition we take the incoming gravitons to possess
momentum in the $t,x$ plane only.
Working in the center of mass frame, and focusing on the case where
the outgoing strings are created
nearly on threshold, it is convenient to parameterize the momenta as
\begin{subequations}
\begin{align}
k_+ &= (m_+, 0, 0) + 
   ( \delta \omega^+, \delta k_x, \delta {\vec k}_T), \\
k_- &= (m_-, 0, 0) + 
   ( \delta \omega^-, - \delta k_x, -\delta {\vec k}_T), \\ 
k_L &= (-\omega, -\omega, 0) + 
  ( - \delta \omega^L, - \delta k'_x, 0 ), \\
k_R &= (-\omega, +\omega, 0) + 
   ( - \delta \omega^R, + \delta k'_x, 0 ),
\end{align}
\end{subequations}
Momentum conservation and the mass--shell conditions imply that 
that we may choose 
 $\delta k_x$ and 
$\delta {\vec k}_T$ to be the independent variables in the problem.  

For our discussion, we will find it useful to define
\begin{subequations}\label{eq:def_ab}
\begin{align}
a &= N_+ - \frac{\alpha' t}{4}, \\
b &= N_- - \frac{\alpha' u}{4},
\end{align}
\end{subequations}
where $t$ and $u$ are the conventional Mandelstam variables.  For the problem
at hand, when the strings are produced with zero spatial
momenta, these variables are given by
\begin{subequations}\label{eq:mandelstam}
\begin{align}
s &= m_+^2 + m_-^2 + 2 m_+ m_-, \\
t &= -m_+ m_-, \\
u &= -m_+ m_-.
\end{align}
\end{subequations}
It will be important for our argument below that in the case where the
pair is produced with zero spatial momentum, these variables are all
$4/\alpha'$ multiplied by integers.

With this parameterization of the momenta, standard techniques in string
perturbation theory lead to the leading order term in the string $S$--matrix,
given by
\begin{multline}\label{eq:str_pp_amp}
\epsilon^+_{\mu_1 {\bar \mu}_1 \cdots \mu_{N_+} {\bar \mu}_{N_+} }
\epsilon^-_{\nu_1 {\bar \nu}_1 \cdots \nu_{N_-} {\bar \nu}_{N_-} }
\epsilon^L_{\sigma {\bar \sigma}} \epsilon^R_{\tau {\bar \tau}}
(N_+!)^{-1} (N_-!)^{-1}
\times \\
\frac{16\pi i}{\alpha'} g_c^2
\left(\frac{\alpha'}{2}\right)^{N_+ + N_-}
\prod_{j=1}^{N_+} k_{R}^{\mu_j} k_R^{{\bar \mu}_j}
\prod_{k=1}^{N_-} k_{R}^{\nu_j} k_R^{{\bar \nu}_j}
\eta^{\sigma\tau} \eta^{{\bar \sigma} {\bar \tau}} \times \\
 \left[ 
\frac{\Gamma(a)\Gamma(b)}{\Gamma(a+b)} \right]^2
\frac{ \sin (\pi a) \sin (\pi b) }{ \sin [ \pi ( a+b) ]}.
\end{multline}
This amplitude is then multiplied by the metric perturbations and integrated according to equation \er{eq:str_prescription}.
There are two key features of this amplitude that differ substantially
from the field theory case.

The first remarkable feature of this amplitude is that the production of
identical string pairs vanishes at zero spatial momentum.  This follows 
directly from the behavior of the sine functions appearing in the
amplitude \er{eq:str_pp_amp}.   When $N_+ = N_- = N$, and when the 
spatial momentum of the created pair vanishes, then using
(\ref{eq:def_ab},\ref{eq:mandelstam}) and the string mass shell condition
we find
\be
a = b = 2N - 1,
\ee
and therefore the trigonometric factors vanish.
For field theory, examination of the 
$S$--matrix element arising from 
\er{eq:qft_2pt_ex} reveals that the pair production amplitude is independent
of $k_{x,T}$ in the long--wavelength limit.  For strings, by contrast,
it appears that the pair production vanishes as a power law in the
long--wavelength limit.  Similar behavior obtains whenever the condition
$a,b \in \mathbf{Z}$ is satisfied, which occurs for pairs of excitation
levels such as
\be
(N_+,N_-) = ( 2, 5), (2, 10), (2, 17), (3, 9), \dots
\ee
in addition to the $N_+ = N_-$ cases corresponding to the creation of
identical pairs.  When $N_\pm$ are such that
$a \notin \mathbf{Z}$ and $b \notin \mathbf{Z}$, then the production
of these string pairs are independent of spatial momentum in the 
long wavelength limit, similar to the field theory case.

This behavior is a consequence of some fundamental features of string theory.
Recall that the open string Veneziano amplitude is essentially uniquely 
determined by
the presence of poles corresponding to excited string states, as well as
worldsheet duality.  The closed string amplitudes are in a sense products
of open string amplitudes, with additional sine factors to ensure that all of 
poles that occur in the product are simple ones.  Again, the closed
string amplitudes
are essentially uniquely determined by basic features of the theory.
Thus different sine factors
are zero when the $s$, $t$ or $u$ Mandelstam variables correspond to a
physical string state and compensate for some of the poles arising from the
gamma functions in the amplitude.  In our case, when the string pairs are 
produced with zero spatial momentum, the $s$ Mandelstam variable corresponds
to an on--shell string state, but the $t$ and $u$ variables are the negative
of on--shell values.  Thus the sine functions have zeros, but now the gamma
functions have no poles, and so the amplitude vanishes.
Of course, this argument requires that we establish exactly some additional
integers appearing in the arguments to the gamma functions, and this is
done in Appendix \ref{appendix}.  The vanishing of this amplitude holds not only for the
leading term in the $S$--matrix, but for terms of all orders in $k$ appearing
in this amplitude.

The second key difference between the string and field theory pair production
amplitudes is an exponential suppression of the string pair production
rate.  
To explore this feature we consider the case where excitation level of
both strings is approximately $N$, in which case $a\sim b \sim 2N$ from
(50). Applying Stirling's approximation to the gamma function factors in
(49) we find

\be
\frac{\Gamma(2N)^4}{\Gamma(4N)^2} =
\frac{2\pi}{N} 2^{-8N}
\left(1 + \frac{1}{8N} + \cdots \right)
\qquad
N \gg 1.
\ee

Therefore, the string amplitude falls off exponentially with $m^2$, in 
contrast to the polynomial dependence of the field theory result. 
This behavior is characteristic of hard scattering string processes.
As we have remarked earlier, typically we expect pair production rates 
in field theory
to fall off exponentially with mass.  However, recall that when studied with
the $S$--matrix approach employed herein, there are actually two effects at
play.  The first is the structure of the field theory amplitude connecting
the incoming gravitons to the outgoing particles.  The second is the 
number of gravitons, or Fourier coefficients of the metric perturbations.  
In field theory, the former behaves polynomially with mass, while the second
drops off exponentially.  For strings on the same background (that is, with the
same Fourier coefficients for metric perturbations) both factors fall off 
exponentially.  Thus fewer pairs are produced.

Again, this may be understood using simple string physics.  
In studies of string amplitudes, one of the kinematic regions of interest
\cite{Gross:1987ar,Gross:1987kz,Gross:1988ue}
is the so--called ``hard scattering" limit
\begin{equation}
s \to \infty, \qquad t/s \quad \mbox{fixed}.
\end{equation}
This is precisely the limit of interest when we look at particles of 
the same mass and look at the limit in which $m \to \infty$.  In the
hard scattering limit, string amplitudes behave as
\begin{equation}
\mbox{amplitude} \sim e^{-s f(\theta)}
\end{equation}
where $f(\theta) > 0$.  This is precisely the behavior observed in
our results.  It indicates that, just as the scattering of strings is softer
than that of point particles at high energy, so too the pair production of
strings is reduced.

\section{Conclusions}\label{s:s6}

In this work we have studied the pair production of excited strings in a
time--dependent background.  The specific background we consider consists
of two colliding plane gravitational waves.  Through studying the singularity
structure of this background, as well as the nature of corrections needed
to cancel ``tadpole" diagrams, we have concluded that reliable results may be 
obtained through standard $S$--matrix methods at second order in the
gravitational wave amplitude.  Our calculations revealed two essential
differences between the field and string theory results.  The pair production
of identical string states is suppressed for certain pairs of excitation 
numbers of the outgoing strings; most significantly, the production of
identical string states vanishes in the infinite wavelength limit.  In addition,
the overall production of strings is suppressed relative to point--particle
results, which may be viewed as a consequence of the mild hard scattering
behavior of string amplitudes.

Our results are
especially relevant to questions regarding
the pair production of strings in cosmological spacetimes.  
In particular,
these results may have implications for one motivation for the current work,
in which the pair production of strings is studied using an effective field
theory approximation \cite{Gubser:2003vk,Gubser:2003hm,Friess:2004zk}.
The suppression of production of excited string states suggests that, 
despite the exponentially growing Hagedorn density of states, effective
field theory provides an overestimate of the production of individual string
states.

Of course, we have established these results using techniques that are
reliable only when spacetime is nearly Minkowski.  Nevertheless, our findings
suggest phenomena that may also be present in the strong
field regime.  It would be quite interesting to know if these effects persist
in more highly curved, time--dependent backgrounds.  If so, 
it would be an example of a uniquely ``stringy" signature of relevance
to cosmology, providing further
clues to the role of quantum gravity in the early universe.

\begin{acknowledgments}
We would like to thank 
Steven Gubser, Justin Khoury, Paul Steinhardt and Herman Verlinde
for useful
comments and discussions.  This work was supported in part by
US Department of Energy Grant DE-FG02-91ER40671 (AJT) and an NSF Graduate
Research Fellowship (DHW).
\end{acknowledgments}

\appendix

\section{$gg \to NM$}\label{appendix}

Standard techniques in string perturbation theory give the form of the
four--point string $S$--matrix amplitude as
\begin{equation}
\mathcal{M} = 
\frac{8 \pi i}{\alpha' g_c^2}
\langle
: c(z_+) {\tilde c}(z_+) \vtxc{k_+}{z_+}{+} :
: c(z_-) {\tilde c}(z_-) \vtxc{k_-}{z_-}{-} :
: c(z_L) {\tilde c}(z_L) \vtxc{k_L}{z_L}{L} :
: \vtxc{k_L}{z_R}{R} :
\rangle
\end{equation}
where the fact that this expression is to be integrated over the unfixed
vertex operator position is understood.  We have also
suppressed the momentum conservation factor, so that
the full $S$--matrix is
\begin{equation}
S_4(k_+,k_-,k_L,k_R) = (2\pi)^{26} \delta^{26} (k_+ + k_- + k_L + k_R)
\mathcal{M} 
\end{equation}
In this work, we take $\vtxc{k_+}{z_+}{+}$ and
$\vtxc{k_-}{z_-}{-}$ to be representative states at level $N_+$ and
$N_-$, respectively, and $\vtxc{k_L}{z_L}{L}$ and
$\vtxc{k_L}{z_R}{R}$ to be the massless graviton states.  We fix the
three
vertex operator positions $z_+ =0$, $z_- = 1$ and $z_L = \infty$.
Next, using momentum conservation and the transverse property of the
polarization tensors of each state, it is possible to put the
matrix element in the form
\begin{multline}
\Bigg\langle
:\prod_{j=1}^{N_+} \left( \partial X^{\mu_j} + \frac{i\alpha'}{2}
\left[ \kappa^{\mu_j} + \frac{k_R^{\mu_j}}{z} \right] \right) :
: \prod_{k=1}^{N_-} \left( \partial X^{\nu_k} - \frac{i\alpha'}{2}
\left[ \kappa^{\nu_k} + \frac{k_R^{\nu_k}}{1-z} \right] \right) : \times \\
: \left( \partial X^\sigma + \frac{i\alpha'}{2} 
\left[ z \kappa^\sigma - k_-^\sigma \right] \right) :
: \left( \partial X^\tau - \frac{i\alpha'}{2}
\left[ \frac{\kappa^\tau}{z} - \frac{k_-^\tau}{z(1-z)} \right] \right) :
\Bigg\rangle
\times \\
 z^{\alpha' k_+ \cdot k_R / 2}
(1-z)^{\alpha' k_- \cdot k_R / 2}
.
\end{multline}
multiplied by its complex conjugate from the antiholomorphic sector,
multiplied by an overall factor
\begin{equation}
\frac{8\pi i g_c^2}{\alpha'} 
\left(\frac{2}{\alpha'}\right)^{N+M+2}
\end{equation}
and of course the appropriate polarization tensor for each state.  In
this expression we have defined, $\kappa^\sigma = k_+^\sigma + k_-^\sigma$.
This expression for the amplitude results from evaluating the ghost path
integral and contractions between the $\partial X$ and $e^{ikX}$ terms, as
well as between two $e^{ikX}$ terms.  The remaining contractions are
those between two $\partial X$ terms.  

Given this pattern of contractions, each possible contraction of operators
results in an integral of the form
\begin{multline}\label{eq:zint}
\int_{\mathbf{C}} z^{a-1+m_1} {\bar z}^{a-1+n_1} 
(1-z)^{b-1+m_2}(1-\bar{z})^{b-1+n_2} \; d^2 z \\
= 2\pi
\frac{\Gamma(a+m_1)\Gamma(b+m_2)\Gamma(1-a-b-n_1-n_2)}{
\Gamma(a+b+m_1+m_2)\Gamma(1-a-n_1)\Gamma(1-b-n_2)}
\end{multline}
For $m_1, m_2, n_1, n_2 \in \mathbf{Z}$.  We can simplify our discussion by
focusing on the leading terms in this expression.  Our concern in the 
present work is on the production of long--wavelength, highly excited
string states.  Thus, we parameterize the momenta in the problem as
\begin{subequations}
\begin{align}
k_+ &= (m_+, 0, 0) + 
   ( \delta \omega^+, \delta k^+_x, \delta {\vec k}_T^+), \\
k_- &= (m_-, 0, 0) + 
   ( \delta \omega^-, \delta k^-_x, \delta {\vec k}_T^-), \\ 
k_L &= (-\omega, -\omega, 0) + 
  ( - \delta \omega^L, - \delta k_x^L, - \delta {\vec k}_T^L ), \\
k_R &= (-\omega, +\omega, 0) + 
   ( - \delta \omega^R, - \delta k_x^R, - \delta {\vec k}_T^R ),
\end{align}
\end{subequations}
where we assume that $\delta k, \delta\omega \ll m_\pm$.  We focus on the
case where
\begin{subequations}
\begin{align}
0 &= \delta {\vec k}_T^L = \delta {\vec k}_T^R, \\
\delta k_x &= \delta k_x^+ = -\delta k_x^-,
\end{align}
\end{subequations}
Now applying the mass--shell conditions and momentum conservation, we find, $\omega = \frac{m_+ + m_-}{2}$
and furthermore,
$\delta {\vec k}_T = \delta {\vec k}_T^+ = - \delta {\vec k}_T^- $, and 
$\delta k_x' = \delta k_x^L = - \delta k_x^R$.
We can take the two remaining free variables to 
be $\delta k_x$ and $\delta {\vec k}_T$. The mass--shell conditions further
imply,
\be
\delta \omega^\pm = 
    \frac{ (\delta k_x)^2 + (\delta {\vec k}_T)^2 }{2m_\pm}
    + \mathcal{O} (k_{x,T}^4), \\
\ee
and $\delta \omega^L = \delta k_x'$, $\delta \omega^R = \delta k_x'$.
So now we can reexpress all variables in terms of $\delta k_x$ and
$\delta {\vec k}_T$, by noting, $\delta k_x' = \frac{\delta \omega^+ + \delta \omega^-}{2}$.
Now returning to the expression \er{eq:zint}, we can see that
\begin{subequations}
\begin{align}
a-1 = (\alpha'/2) k_+ \cdot k_R = N_+ - 1 + \Delta + \delta_{+R}, \\
b-1 = (\alpha'/2) k_- \cdot k_R = N_- - 1 + \Delta + \delta_{-R},
\end{align}
\end{subequations}
with $\Delta = \left[ (N_+ - 1)(N_- - 1) \right]^{1/2}$
and,
\begin{subequations}
\begin{align}
\delta_{-R} & = \frac{\alpha'}{2} \left( \frac{m_-}{2} \delta \omega^+
  + \left[\frac{m_+ + 2m_-}{2}\right] \delta \omega^- 
  - \omega \delta k_x \right) ,\\
\delta_{+R} & = \frac{\alpha'}{2} \left( \frac{m_+}{2} \delta \omega^-
  + \left[\frac{m_- + 2m_+}{2}\right] \delta \omega^+ 
  + \omega \delta k_x \right), 
\end{align}
\end{subequations}
At this point we have all of the information we need to compute each term in the
amplitude.  In order to find the dominant term, it suffices to examine each
of the possible contractions between momentum vectors and polarization tensors.
These are given by, $(k_R \cdot e^+,k_R \cdot e^- )=\mathcal{O}(\sqrt{\alpha'} m)$ and $(\kappa \cdot e^+,\kappa \cdot e^-,\kappa \cdot e^L,k_- \cdot e^L,\kappa \cdot e^R,k_- \cdot e^R)=\mathcal{O}(\sqrt{\alpha'} \delta k)$.
%
%
%
Thus it is seen that the term with the greatest number of $k_R$ contracted
with polarization tensors will be dominant.  It will be convenient to 
rewrite the integral \er{eq:zint} as
\begin{multline}
\pi
\frac{\Gamma(a+m_1)\Gamma(b+m_2)\Gamma(1-a-b-n_1-n_2)}{
\Gamma(a+b+m_1+m_2)\Gamma(1-a-n_1)\Gamma(1-b-n_2)} = \\
 \frac{\sin [\pi (a+n_1)] \sin [ \pi(b+n_2)]}{\sin [ \pi ( a+b+n_1+n_2) ]}
\frac{\Gamma(a+m_1)\Gamma(b+m_2)}{\Gamma(a+b+m_1+m_2)}
\frac{\Gamma(a+n_1)\Gamma(b+n_2)}{\Gamma(a+b+n_1+n_2)}
\end{multline}
From our discussion above, the leading term with have the largest possible
number of $k_R$ contractions, implying $m_1 = n_1 = -N_+$ and
$m_2 = n_2 = -N_-$.  (This depends somewhat on the pattern of contractions
involving the graviton polarization tensors, but this turns out to be
unimportant in the final result)  Thus we find the value of the integral
to be
\begin{equation}\label{eq:amp_factor}
  \frac{\sin [\pi (\Delta + \delta_{+R})] \sin [ \pi(\Delta +
\delta_{-R})]}{\sin [ \pi ( 2\Delta + \delta_{+R} + \delta_{-R}) ]}
\left[ 
\frac{\Gamma(\Delta + \delta_{+R}) \Gamma( \Delta + \delta_{-R})}{\Gamma(2\Delta
+ \delta_{+R} + \delta_{-R})}
\right]^2 
\end{equation}
%
%
From this it is possible to work backwards and obtain the full $S$--matrix
expression, although we will not require the full amplitude.

A key feature of this amplitude is its behavior when $\Delta \in \mathbf{Z}$.
The leading term derived above clearly vanishes as $\delta_{\pm R} \to 0$
in this case, due to the
structure of the zeros of the sine functions.  While we have only displayed
the leading order term above, this behavior in fact holds for all of the
terms contributing to this amplitude.  This can be seen from the structure
of the integral over the complex plane used to derive \er{eq:amp_factor}.
Regardless of the pattern of contractions corresponding to a given term,
the arguments to all of the gamma functions appearing in 
\er{eq:amp_factor} are positive.  Since the gamma function has no zeros, and
only has poles for negative values of the argument, the amplitude must
vanish when $\Delta \in \mathbf{Z}$ and $\delta_{\pm R} \to 0$.  

\section{Pair Production via the $S$--matrix}\label{appendix2}

In this appendix we carry out explicit calculations of the pair 
production rate using both $S$--matrix and Bogolubov coefficient techniques.
This provides a check that, while the two methods differ, the end result
is precisely the same (for a related analysis, see also \cite{Hamilton:2003xr}).
We study a minimally coupled scalar field in $D \ne 2$ spacetime 
dimensions, with action
\begin{equation}
\mc{S}_\phi = -\frac{1}{2} \int \left( g^{\mu\nu} \partial_\mu \phi \partial_\nu \phi
+ m^2 \phi^2 \right) \sqrt{-g} \; d^D x,
\end{equation}
Our gravitational background is defined by
\begin{equation}
g_{\mu\nu} = a(t)^2 \eta_{\mu\nu}
\end{equation}
where the ``scale factor" $a(t)$ is an arbitrary function of time.  In order to
better make contact with the examples considered in this work, we will assume that
$a(t)$ is unity, except for a localized ``bump" of small amplitude near
$t=0$.  Thus, there will be well--defined ``in" and ``out" regions where the
geometry is Minkowski space.

For future convenience we define a field variable
$\psi$, given by
\begin{equation}
\phi(t,x_j) = a (t)^{1 - D/2} \psi (t,x_j),
\end{equation}
We Fourier expand $\psi$ along the spatial coordinates, so
$\psi(t,x_j) = \psi_k(t) e^{ik_j x^j}$, and define $g(t) = a^{1/n}$, $n = \frac{1}{D/2 - 1}$,
and
\begin{equation}
\Delta(t) = \left( g(t)^{2n} - 1 \right) m^2 - \frac{\ddot g}{g},
\end{equation}
With these replacements, the action for
$\psi$ becomes
\begin{equation}
\mc{S}_\psi = -\frac{1}{2} \int \left( 
- {\dot \psi_k}^2 +  \omega_k^2 \psi_k^2
+ \Delta (t) \psi_k^2
\right) \; \frac{d^{D-1} k}{(2\pi)^{D-1}} dt ,
\end{equation}
where $\omega_k^2 = m^2 + k^2$ as usual.
The equation of motion for $\psi_k$ is
\begin{equation}
\ddot \psi + \left[
  \omega_k^2 + \Delta(t)
  \right] \psi = 0.
\end{equation}

\subsection{Bogolubov Coefficient Calculation}

Now, we will calculate precisely what the $\beta$ Bogolubov coefficient is in
the theory described in the previous section.  Because the perturbation to the
action is localized in time, the canonical Minkowski vacuum is a natural choice
for the ``in" and ``out" regions, and so
\begin{equation}
u_j(t,x) = {\bar u}_j(t,x) = e^{-i\omega t + ik_j x^j}, \qquad
u_j^*(t,x) = {\bar u}^*_j(t,x) = e^{+i\omega t - ik_j x^j},
\end{equation}
We now consider the following (approximate) solution to the
equation of motion for $\psi$
\begin{equation}\label{eq:perturb_soln}
\psi_k(t) = \frac{1}{\sqrt{2\omega(k)}} \left(
 e^{-i\omega_k t} +
 \beta_{k,-k'} (t) e^{i\omega_k t} \right)
\end{equation}
along with the boundary condition that $\beta_{k,-k}(t) \to 0$ as
$t \to -\infty$.  This boundary condition ensures that $\psi_k(t)$ is a pure
positive--frequency mode in the ``in" region.  Using the definitions above, it
is clear that $\beta_{k,-k}(t)$ at $t = +\infty$ is the Bogolubov $\beta_{k,-k}$
coefficient.  
  
Substituting our solution \er{eq:perturb_soln} into the
wave equation, and making the assumption that
$\beta \ll 1$, 
we find
\begin{equation}
\ddot \beta_{k,-k} + 2i\omega_k \dot \beta_{k,-k} = 
- \Delta(t) e^{-2i\omega_k
t}
\end{equation}
This equation has the solution
\begin{equation}
\beta_{k,-k}(t) = - \int_{t' = -\infty}^{t} e^{-2i\omega_k t'}
\int_{t'' =  -\infty}^{t'} \Delta(t'') \; dt' \, dt''
\to
\frac{i}{2\omega_k} \Delta (2\omega_k),
\end{equation}
where in the last step we have taken $t \to +\infty$.  Note that our derivation
is self--consistent even in $\Delta(t)$ is large; we have only assumed that
$\beta_k(t) \ll 1$ and thus we only need require that 
$\Delta(\omega)/\omega$ is small.

To compare with the $S$--matrix calculation, we must switch to the 
Lorentz--invariant particle states, defined by
\be
|k\rangle = \sqrt{2\omega_k} a^+_k |0\rangle .
\ee
When this is done, one
arrives at the invariant matrix element describing the particle production
process,
\begin{equation}
\mathcal{M}\left( | 0 \rangle \to |k,-k\rangle \right) = -
\Delta (2\omega_k)
\end{equation}
We will see in the next section how precisely the same result is obtained using
conventional field theory techniques.

\subsection{$S$--matrix Calculation}

The $S$--matrix element may be calculated using the rules of standard
flat--space quantum field theory.  
The calculation is much simpler than using the Bogolubov coefficient method.
In quantum field theory, if we have a perturbation
to the action $\delta S [\psi]$, then
\begin{equation}
S = \int_{vol} \langle 0 | i\delta S [\psi] | k_1 k_2 \rangle
\end{equation}
where we ``contract" any $\psi$ appearing in the perturbation according to the
rule,
\begin{equation}
\psi(x) | k_1 \rangle = e^{-ik_1\cdot x} | 0 \rangle
\end{equation}
In our situation, we have $\delta S = - \Delta(t) \psi^2 / 2$.  Again, after
including the factor of two from summing over the two possible contractions
with the final state momenta, we find that,
\begin{equation}
S = -i \int e^{-i(k_1 + k_2)\cdot x} \Delta(t) \; d^D x,
\end{equation}
which is precisely the Fourier transform of $\Delta(t)$.  Imposing momentum
conservation, we find the invariant matrix element,
\begin{equation}
\mathcal{M}\left( | 0 \rangle \to |k,-k\rangle \right) = -
\Delta (2\omega_k)
\end{equation}
which is precisely the one found by other means in the previous section.

\end{document}